\documentclass[aps,prl,twocolumn,showpacs,superscriptaddress,groupedaddress]{revtex4}
\usepackage{graphicx}
\usepackage{dcolumn}
\usepackage{bm}

\begin{document}


\title{Definition of the stimulated emission threshold in high-$\beta$ nanoscale lasers through phase-space reconstruction}

\author{X.\ Hachair$^{1}$, R.\ Braive$^{1, 2}$, G.L.\ Lippi$^{3,4}$, D.\ Elvira$^{1}$, L.\ Le Gratiet$^{1}$,  A.\ Lemaitre$^{1}$, I.\ Abram$^{1}$, I.\ Sagnes$^{1}$, I.\ Robert-Philip$^{1}$ and A.\ Beveratos$^{1,*}$}

\address{
$^1$CNRS - Laboratoire de Photonique et de Nanostructures, UPR20,
Route de Nozay, 91460 Marcoussis, France
\\
$^2$Universit\'e Paris Diderot-Paris 7 \\
$^3$Université de Nice-Sophia Antipolis, Institut Non Linéaire de Nice\\
$^4$CNRS, UMR 6618\\
$^*$Corresponding author: alexios.beveratos@lpn.cnrs.fr }

\begin{abstract}

Nanoscale lasers sustain few optical modes so that the fraction of spontaneous emission $\beta$ funnelled into the useful (lasing) mode is high (of the order of few 10$^{-1}$) and the threshold, which traditionally corresponds to an abrupt kink in the light in- light out curve, becomes ill-defined. We propose an alternative definition of the threshold, based on the dynamical response of the laser, which is valid even for $\beta=1$ lasers. The laser dynamics is analyzed through a reconstruction of its phase-space trajectory for pulsed excitation. Crossing the threshold brings about a change in the shape of the trajectory and in the area contained in it. An unambiguous definition of the threshold in terms of this change is shown theoretically and illustrated experimentally in a photonic crystal laser.

\end{abstract}

\pacs{42.55.Tv, 42.55.Sa, 42.55.Ah} \maketitle

Recent progress in the fabrication of nanoscale lasers has opened the possibility of developing extremely efficient devices in which even the spontaneously emitted photons are preferentially funneled into the lasing mode. 
In these devices, the fraction $\beta$ of spontaneous emission going into the laser mode becomes a significant fraction of unity, in contrast to conventional lasers, in which $\beta \simeq 10^{-5}$. 
High-$\beta$ nanolasers may be obtained in high-Q small-volume cavities, such as micropillars \cite{Ulrich2007,Wiersig2009} or photonic crystal nanocavities \cite{Nomura2007,Hostein2010,Choi2007}, where the funneling is due to the directional enhancement of spontaneous emission through the Purcell effect, nanowires \cite{Oulton2009}, in which the high $\beta$ results from the very small number of modes they sustain \cite{Lecamp2007}, or surface plasmon lasers (SPASERs) \cite{Noginov2009,Hill2009,Nezhad2010}, where the plasmon field strongly enhances spontaneous emission. High-$\beta$ nanolasers can also be obtained in the strong coupling regime, using polariton condensation \cite{Assmann2009}. 

When $\beta$ approaches unity, the transition from spontaneous to stimulated emission becomes smooth, so that the traditional definition of the threshold as an abrupt change in the Light-in Light-out ($L-L$) curve of the device is no longer applicable \cite{Yamamoto1991}. 
In fact, for $\beta=1$ the $L-L$ curve is a perfect straight line, with no break, so that from the viewpoint of energy efficiency the $\beta=1$ laser can be considered ``thresholdless'' \cite{Yokoyama1992} as it exhibits the same efficiency for both spontaneous and stimulated emission. However, from the viewpoint of dynamical response it is important to be able to distinguish between the spontaneous and stimulated regimes, as the large-signal direct modulation (gain-switching) bandwidth of the laser is different in these two regimes. In addition, in high-$\beta$ lasers, thanks to the deterministic channelling of spontaneous photons into the lasing mode, stimulated emission is initiated very rapidly giving rise to a broad direct modulation bandwidth even for large amplitude modulation (On-Off Keying -- OOK) , a feature desirable for high-speed optical data transmission.

In order to differentiate the spontaneous from the stimulated regime, it was proposed to monitor the threshold through the evolution of coherence, in particular through the increase in the first order coherence, as dictated by the Schawlow-Townes relation \cite{Ates2008,Nomura2007} (which, however, can only be measured for stabilized continuous-wave operation) or through the decrease of the second order autocorrelation function, from g$^{(2)}(0)=2$ below threshold (corresponding to chaotic light) to g$^{(2)}(0)=1$ (corresponding to coherent emission) \cite{Ulrich2007,Choi2007,Hostein2010,Assmann2009,Wiersig2009}. However both techniques have fundamental limitations. In nanoscale lasers, thermal or refractive index fluctuations lead to a deviation from the Schawlow-Townes expectations, while the transition from g$^{(2)}(0)=2$ to unity is not abrupt \cite{Ulrich2007} and thus does not permit a more precise determination of the threshold than the $L-L$ curve. 
At the same time, the threshold obtainable from the evolution of coherence is not related in a simple way to the stimulated emission threshold and does not provide direct information on the response dynamics of the laser.
In addition, a definition of the threshold based on g$^{(2)}$ and the underlying amplitude fluctuations would be inapplicable to lasers displaying non-classical photon statistics, such as amplitude squeezed lasers \cite{Machida1987}. 

In this Letter, we propose a phase-space reconstruction technique that permits an unequivocal identification of stimulated emission and provides an unambiguous definition of the threshold in nanolasers that can also be used for nanolasers with $\beta=1$. We illustrate this approach through experiments identifying spontaneous and stimulated emission in the output of a high-$\beta$ ($\beta \approx 0.25$) nanolaser.

Our analysis is based on the traditional laser rate equations which, for a simple laser, are \cite{Druten2000}:
\begin{eqnarray}
\label{eq:Rate1}
\frac{dN(t)}{dt} = & P f(t) -\gamma_{\parallel}N(t) & -\beta \gamma_{\parallel} N(t) n(t) \\ 
\label{eq:Rate2}
\frac{dn(t)}{dt} = & -\Gamma_c n(t) +\beta \gamma_{\parallel} N(t) &+ \beta \gamma_{\parallel} N(t) n(t)
\end{eqnarray}
where $N(t)$ is the number of excited dipoles, $n(t)$ the number of photons in the cavity, $\beta$ the ratio of spontaneous photons in the laser mode over the total number of spontaneously emitted photons, $\gamma_{\parallel}^{-1}$ the dipole lifetime, $\Gamma_c$ the cavity loss rate and $P f(t)$ describes the pumping, with the integral of $f(t)$ over time, equal to 1. 
It should be noted that, because of the short round-trip time of light in the nanoscale cavity, the cavity decay is fast, so that nanolasers are generally characterized by $\Gamma_c \geq \gamma_{\parallel}$ (Class-B lasers).
Following \cite{Loudon2003}, we can define $N_c=\frac{\Gamma_c}{\beta \gamma_\parallel}$ the critical number of dipoles necessary to attain the threshold.
More complicated rate equations, taking into account various effects of the gain material, such as absorption of ground-state dipoles or non-radiative losses, would not alter the general discussion of our approach. 

Our analysis requires that the system be unexcited before and after the experiment, in other words that the pumping last only for a finite time, so that the number of photons and excited dipoles return to zero after the pumping is over.
To visualize the evolution of the laser under Eqs.\ (\ref{eq:Rate1}) and (\ref{eq:Rate2}), we represent its response in a two-dimensional phase-space defined by $\{dn(t)/dt,n(t)\}$ \cite{Packard1980}, in a manner analogous to that for the traditional representation of transient laser dynamics through the parametric curve $\{N(t),n(t)\}$ in which time is factored out \cite{Lippi2000}.
The advantage of this novel phase-space is that, in contrast to $N(t)$ which cannot be measured directly, $dn(t)/dt$ like  $n(t)$ are both accessible experimentally, through the signal intensity measured outside the cavity, $S(t)$ and $dS(t)/dt$.
As the system returns to its initial (ground) state at the end of the excitation process, its trajectory in phase-space is a closed curve.
\begin{figure}[h!]
   \begin{center}
   \begin{tabular}{c}
   \includegraphics[width=7.5cm]{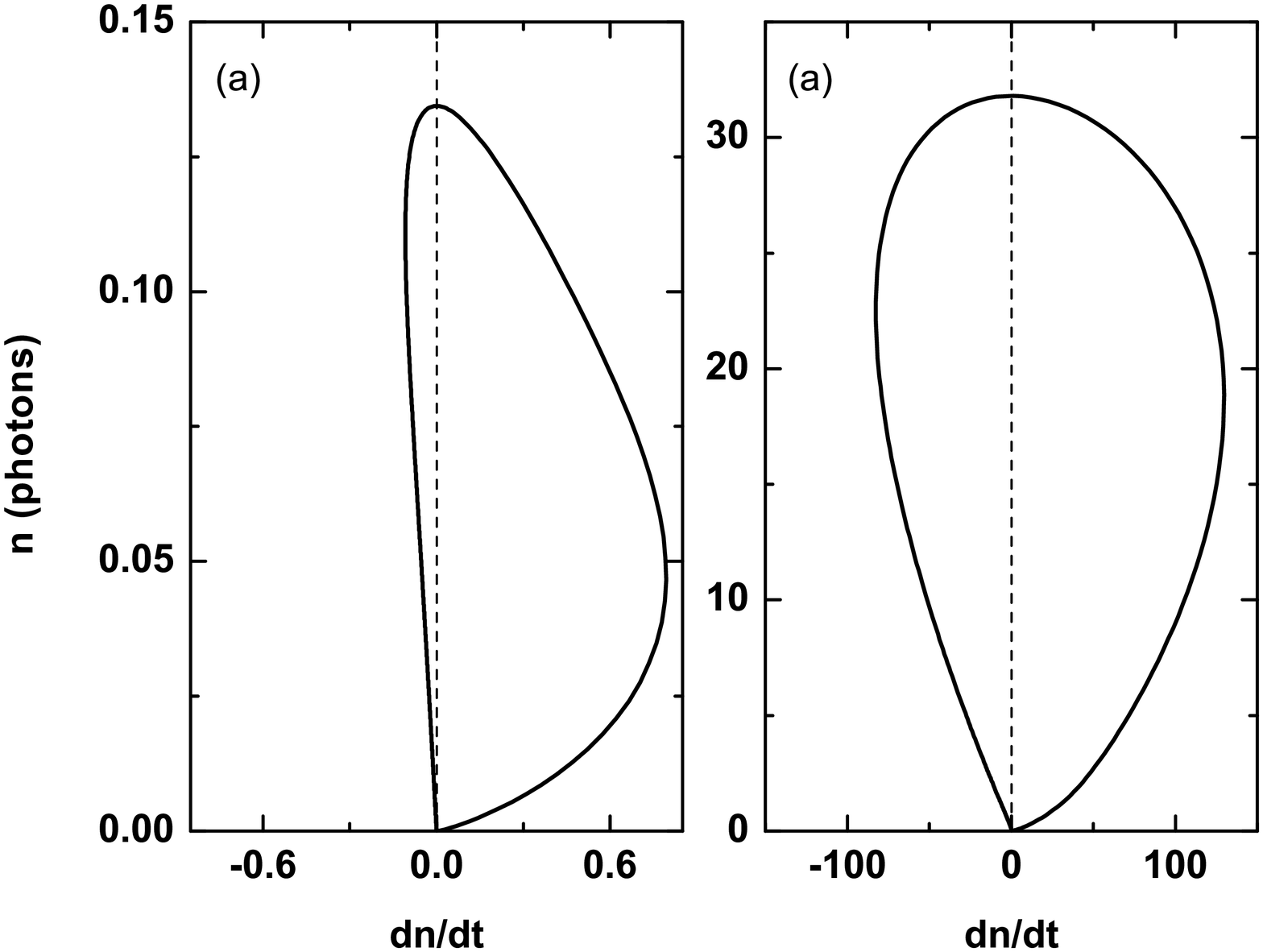}\\
   \includegraphics[width=8cm]{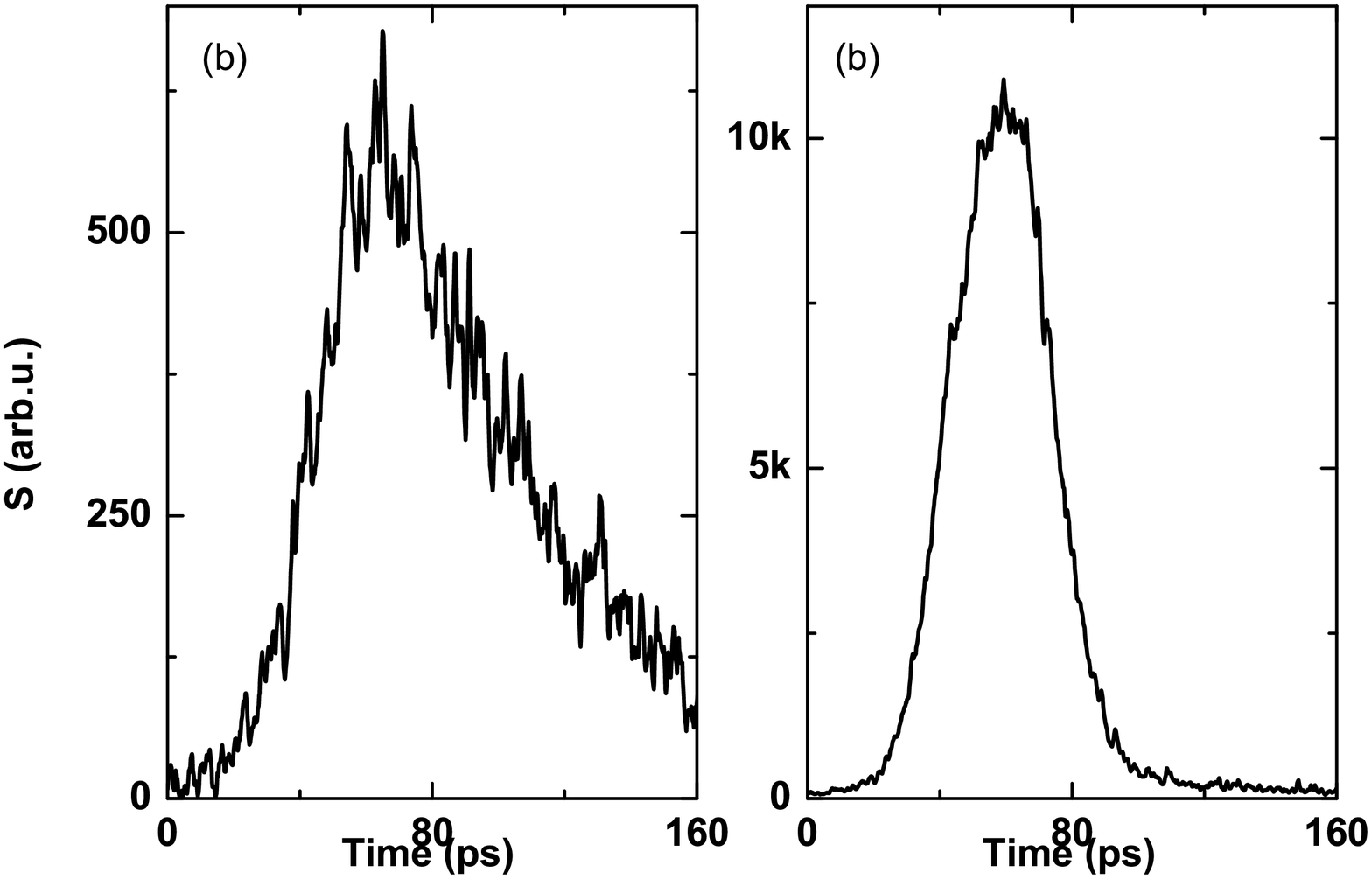}\\
   \includegraphics[width=8cm]{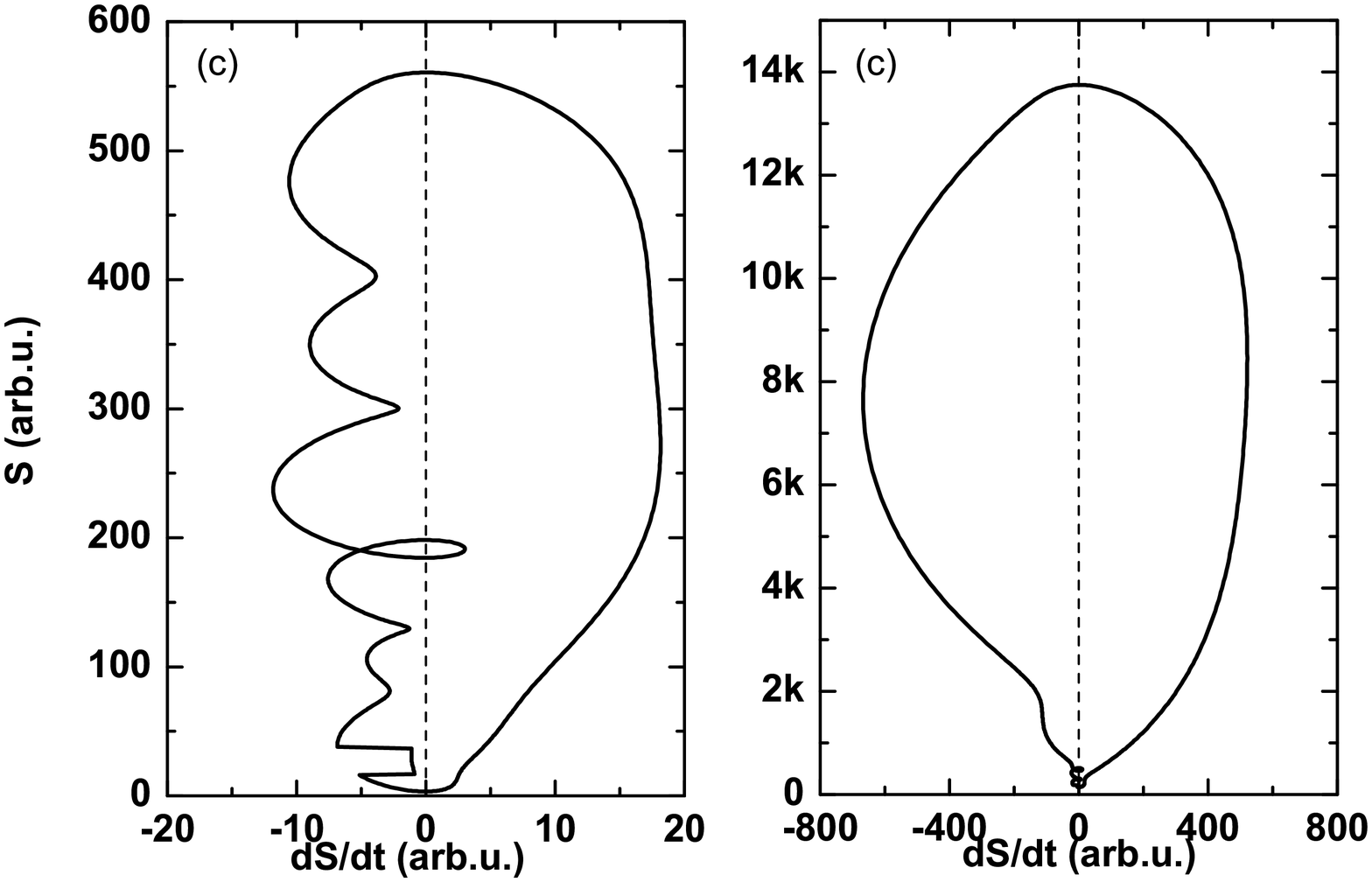}
   \end{tabular}
   \end{center}
   \caption[example]
   {Phase-space trajectories of the nanolaser response: 
   (a) Theoretical trajectories for a nanolaser of $\beta=0.1$ and $\Gamma_c=10\gamma_{\parallel}$, under pumping  below ($P= 0.1 N_c$) and above ($P = 2 N_c$) threshold. 
  (b) Experimental time-traces of the output of a nanolaser under impulse excitation, below ($P_{sp}= 1.7 \mu$J cm$^{-2}$) and above ($P_{st}= 8.6 \mu$J cm$^{-2}$) threshold. 
  (c) Phase-space trajectories corresponding to the experimental time-traces above. }
\label{Fig:loop}
\end{figure} 
The rate equations (\ref{eq:Rate1}) and (\ref{eq:Rate2}) are solved numerically, taking $f(t)$ as a Gaussian pulse much shorter than $\gamma^{-1}_{\parallel}$.
Figure (\ref{Fig:loop}a) represents typical phase-space diagrams below ($P= 0.1 N_c$) and above ($P = 2 N_c$) threshold, for a nanolaser with $\beta=0.1$ and $\Gamma_c=10\gamma_{\parallel}$. 
At low pumping energies, when the laser operates in the spontaneous emission regime, the number of photons $n(t)$ and its derivative $dn(t)/dt$ both increase, but attain relatively small values overall. 
The phase-space trajectory is thus a relatively small loop.
On the other hand, at high pumping energies, when the system operates under full laser dynamics \cite{Hachair2005}, the number of photons $n(t)$ and its derivative $dn(t)/dt$ both increase to high values, so that the area enclosed by the trajectory increases dramatically.
In addition, for Class-B lasers ($\Gamma_c \geq \gamma_{\parallel}$) the shape of the trajectory loop changes as soon as stimulated emission becomes the dominant dynamical process. Below threshold, after $n(t)$ reaches its maximum, the phase space trajectory returns to the origin through the negative half-plane ($dn(t)/dt<0$) quasi-vertically so that the overall curve lies mostly in the right-hand half-plane defined by $dn(t)/dt>0$. In that case $-\Gamma_c n(t)$ is slightly bigger than $\beta \gamma_\parallel N(t) (1+n(t))$. The number of excited dipoles is still large and decays with a characteristic lifetime $\gamma_\parallel$. On the other hand, above threshold, the number of excited dipoles drops considerably due to stimulated emission and the negative term dominates. 
Thus the trajectory makes a long excursion in the negative half-plane so that the phase-space curve becomes almost symmetric. 
Hence, even for $\beta=1$, the effect of stimulated emission is observable in the phase space trajectory and in the area it encloses.

The area enclosed within the phase-space trajectory can be calculated through the Gauss-Green formula for a parametric curve $\{x(t),y(t)\}$:
\begin{equation}
A=\frac{1}{2} \int^{\infty}_{0}{ \left\{ x(t)\frac{dy(t)}{dt}-y(t)\frac{dx(t)}{dt} \right\} dt} ,
\end{equation}
which in our case reads,
\begin{equation}
A=\frac{1}{2} \int^{\infty}_{0}{ \left\{ \left( \frac{dn(t)}{dt} \right) ^2-n(t)\frac{d^2n(t)}{dt^2} \right\} dt} .
\label{eq:PhaseArea}
\end{equation}
Equation (\ref{eq:PhaseArea}) can be integrated by parts, and since at $t=0$ and $t=\infty$ both $n(t)$ and $\frac{dn(t)}{dt}$ are zero under pulsed excitation, the area is given by
\begin{equation}
A=\int^{\infty}_{0}{\left(\frac{dn(t)}{dt}\right)^2 dt}
\end{equation}
which, using Equation (\ref{eq:Rate2}), can be expressed as 
\begin{widetext}
\begin{equation}
A=\int^{\infty}_{0}\Gamma^2_c \left\{ \left(1-\tilde{N}(t) \right)^2 n^2(t)+\tilde{N}^2(t) - 2 \left(1 -\tilde{N}(t) \right)\tilde{N}(t) n(t) \right\} dt
\label{eq:Area}
\end{equation}
\end{widetext}
where $\tilde{N}(t)= N(t)/N_c$.
Below threshold, $\tilde{N}(t)<1$ at all times, so that the third term in Eq.\ (\ref{eq:Area}) is negative and is thus subtracted from the overall phase-space area. On the other hand, this term turns positive above threshold and its contribution adds to the total area during the time it remains positive. 
Thus, the area enclosed by the parametric curve is expected to increase dramatically at the transition between spontaneous and stimulated emission, even in the case $\beta=1$ for which no break in the $L-L$ curve can be observed. 
\begin{figure}[h!]
   \begin{center}
   \begin{tabular}{c}
   \includegraphics[width=9cm]{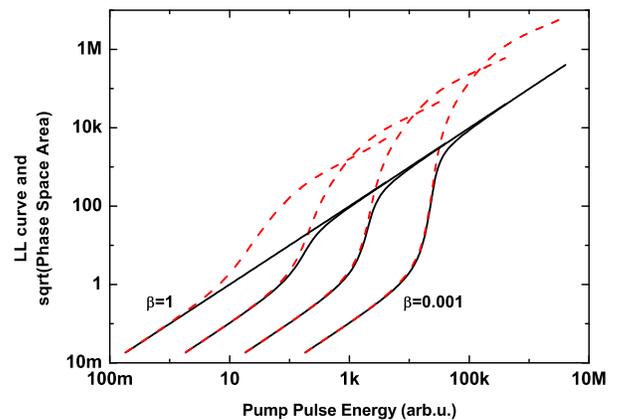}
   \end{tabular}
   \end{center}
   \caption[example]
   { Comparison of the theoretical Light-In Light-Out curves (continuous black lines) and the phase-space area curves (dashed red lines) under pulsed pumping for a nanolaser with $\Gamma_c=10\gamma_{\parallel}$ and $\beta=\{$1, 0.1, 0.01, 0.001$\}$}
\label{Fig:Area}
\end{figure}
Figure \ref{Fig:Area} shows a log-log plot of the calculated $L-L$ curve as well as the square root of the calculated area enclosed by the phase-space trajectory as a function of the pump energy for $\beta=\{$1, 0.1, 0.01, 0.001$\}$ and $\Gamma_c=10\gamma_\parallel$.
The square root is used to facilitate comparison between the two types of curves, since both the signal and the square root of the area are proportional to $P$ below threshold. The two types of curves coincide at low pumping and display the onset of a nonlinear increase (a ``knee") at approximately the same pumping energies. 
However, while the $L-L$ curves display a level change only for $\beta<1$, the phase-space area curves display such a change even for $\beta=1$, permitting identification of stimulated emission in all cases.

In steady-state lasers the threshold is conventionally defined as the $x$-intercept of the high pumping power asymptote in the linear $L-L$ curve, which occurs at $P= (1-\beta) N_c$. 
Clearly, this definition cannot be applied to the $\beta=1$ laser, as it sets the threshold at $P=0$.
Alternatively, the threshold is often defined as the inflection point of the log-log $L-L$ curve, where the number of photons in the cavity per dipole lifetime is $1/\sqrt{\beta}$, so that stimulated emission dominates the response of the laser beyond that point. 
On the other hand, the onset of stimulated emission is associated with the ``knee" of the log-log $L-L$ curve, where the number of photons starts exceeding the value of 1. Thus the "`knee"' of the area curve can be defined as the thershold. This definition is also valid for $\beta=1$ laser and identifies the point beyond which the laser dynamics (including its modulation bandwidth) change.

To illustrate the above considerations, we have investigated experimentally the phase-space transition for a photonic crystal nanolaser. The cavity consists of a photonic crystal double heterostructure \cite{Song2005} etched in a 180 nm-thick suspended GaAs membrane \cite{Braive2009JVST}, incorporating a single layer of self-assembled InAs quantum dots at its vertical center plane. Details of the structure and experimental setup can be found in \cite{Braive2009OL}. 
The nanolaser is optically pumped by a Ti:Sa laser delivering 3 ps-long pulses at a 81.8 MHz repetition rate and tuned to
840 nm near the energy gap of the wetting layer, in order to reduce heating of the membrane. 
Time traces are obtained by a Syncroscan streak camera (Hamamatsu), with a temporal resolution of 3 ps. 
Two representative experimental time traces are represented on Figure \ref{Fig:loop}(b), one in the spontaneous emission regime ($P_{sp}=1.7$ $\mu$Jcm$^{-2}$) and one in the stimulated regime ($P_{st}=8.6$ $\mu$Jcm$^{-2}$). 
The decay rates of these two traces permit us to deduce $\gamma_\parallel^{-1}=50$ ps and $\Gamma_c^{-1}=13$ ps.
The phase-space trajectories were obtained by measuring the cavity signal as function of time $S(t)$ and then computing $dS(t)/dt$ after applying a numerical low pass filter to smooth out the noise in the data. 
Figure \ref{Fig:loop}(c) represents the phase-space trajectories that correspond to the two experimental time-traces of Figure \ref{Fig:loop}(b).
The two key features expected upon passage through the threshold, namely the change in symmetry and the dramatic increase in the area are clearly seen.
\begin{figure}[h!]
   \includegraphics[width=9cm]{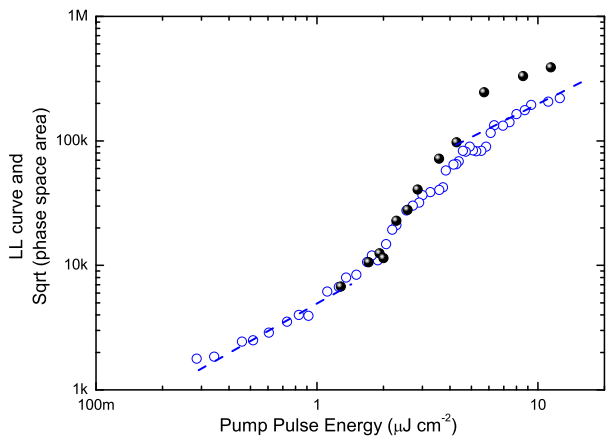}
   \caption[example]
   { \label{Fig2} Log-log plot of the Light in - Light out curve (blue open circles) and the square root of the area enclosed by the phase space trajectory (black circles) as a function of pump pulse energy. The dashed lines of slope unity in log-log scale are a guide to the eye}
   \end{figure}
Figure \ref{Fig2} presents a log-log plot of the square root of the integrated phase-space area, appropriately normalized, and compares it with the $L-L$ curve. 
A spontaneous emission fraction of $\beta \approx 0.25$ can be inferred from the intensity ratio between the spontaneous and stimulated emission levels. 
The ``knee" of the area plot coincides with that of the $L-L$ curve, giving for the threshold $P_{th}=2.1$ $\mu J$cm$^{-2}$.

In summary, the emission properties of nanoscale lasers with large spontaneous emission coupling factors have been
studied from a dynamical point of view. 
In such lasers, the transition from spontaneous to stimulated emission extends over a wide pump power range in the Light in - Light out curve, and the threshold cannot be defined easily, especially when $\beta=1$.
We  provide a simple and unequivocal identification of laser threshold through a new indicator:  the area $A$ enclosed by the phase-space trajectory that corresponds to the impulse response of the device. With this definition a threshold is defined even when none is visible in the $L-L$ curve, as is the case for $\beta \approx 1$ lasers.
This definition is directly related to the onset of stimulated emission and thus directly pertains to the dynamical response of the laser, while at the same time it provides a practical tool for quick and easy measurements.

The authors gratefully acknowledge financial support from the Triangle de la Physique under the BIRD project and from the French National Agency (ANR) through the Nanoscience and Nanotechnology Program (project NATIF ANR-09-NANO-P103-36). The authors also thank K.\ Gauthron and J.\ Bloch for lending critical equipment and S. Barbay for helpful discussions.


\begin{thebibliography}{99}

\bibitem{Ulrich2007} S.M.\ Ulrich, \textit{et. al.}, Phys.\ Rev.\ Lett.\ \textbf{98}, 043906 (2007).

\bibitem{Wiersig2009} J.\ Wiersig, \textit{et. al.}, Nature \textbf{460}, 248 (2009).

\bibitem{Nomura2007} M.\ Nomura, \textit{et. al.}, Phys.\ Rev.\ B \textbf{75}, 195313 (2007).

\bibitem{Hostein2010} R.\ Hostein, \textit{et. al.}, Opt.\ Lett.\ \textbf{8}, 1154 (2010).

\bibitem{Choi2007} Y.-S.\ Choi, \textit{et. al.}, Appl.\ Phys.\ Lett.\ \textbf{91}, 031108 (2007).

\bibitem{Oulton2009} R.F.\ Oulton, \textit{et. al.}, Nature \textbf{461}, 629 (2009).

\bibitem{Lecamp2007} G.\ Lecamp, P.\ Lalanne, J.P.\ Hugonin, Phys.\ Rev.\ Lett.\ \textbf{99}, 023902 (2007).

\bibitem{Noginov2009} M.A.\ Noginov, \textit{et. al.}, Nature \textbf{460}, 1110 (2009).

\bibitem{Hill2009} M.T.\ Hill, \textit{et. al.}, Opt.\ Exp.\ \textbf{17}, 11107 (2009).

\bibitem{Nezhad2010} M.P.\ Nezhad, \textit{et. al.}, Nature Photonics \textbf{4}, 395 (2010).

\bibitem{Assmann2009} M.\ Aßmann, \textit{et. al.}, Science \textbf{325}, 297 (2009).

\bibitem{Yamamoto1991} Y.\ Yamamoto, S.\ Machida and G.\ Bj\"ork, Phys.\ Rev.\ A \textbf{44}, 657 (1991).

\bibitem{Yokoyama1992} H.\ Yokoyama, Science \textbf{256}, 66 (1992).

\bibitem {Ates2008} S.\ Ates, \textit{et. al.}, Phys.\ Rev.\ B \textbf{78}, 155319 (2008).

\bibitem{Machida1987} S.\ Machida, \textit{et. al.}, Phys.\ Rev.\ Lett.\ \textbf{58}, 1000 (1987).


\bibitem{Druten2000} N.J.\ van Druten, \textit{et. al.}, Phys.\ Rev.\ A \textbf{62}, 053808 (2000).

\bibitem{Loudon2003} R.\ Loudon, {\it The Quantum Theory of Light}, Oxford University Press 2003.

\bibitem{Packard1980} N.H.\ Packard, \textit{et. al.}, Phys.\ Rev.\ Lett.\ \textbf{45}, 712 (1980).

\bibitem{Lippi2000} G.L.\ Lippi, \textit{et. al.}, J.\ Opt.\ B Quant.\ Sem.\ Opt.\, \textbf{2}, 375 (2000).

\bibitem{Hachair2005} X.\ Hachair, \textit{et. al.}, Applied Opt.\ \textbf{44}, 4761 (2005).

\bibitem{Song2005} B.-S.\ Song, \textit{et. al.}, Nature Materials \textbf{4}, 207 (2005).

\bibitem{Braive2009JVST}  R.\ Braive, \textit{et. al.}, J.\ Vac.\ Sci.\ Technol.\ B \textbf{27}, 1909-1914 (2009).

\bibitem{Braive2009OL} R.\ Braive, \textit{et. al.}, Opt.\ Lett.\ \textbf{34}, 554 (2009).

\end{thebibliography}
\end{document}